\documentclass[pra,twocolumn,superscriptaddress,showpacs]{revtex4}

\usepackage{graphics}		
\usepackage{bm}			
\usepackage{amsmath}		

\begin{document}

\title[Experimental realization of optimal detection strategies]{Experimental realization of optimal detection strategies for overcomplete states}

\author{Roger B. M. Clarke}
\affiliation{Department of Physics and Applied Physics, University of Strathclyde, Glasgow, G4 0NG, UK}

\author{Vivien M. Kendon}
\altaffiliation[Present address: ]{Optics Section, Blackett Laboratory, Imperial College, London, SW7 2BW, UK, email Viv.Kendon@ic.ac.uk.}
\affiliation{Department of Physics and Applied Physics, University of Strathclyde, Glasgow, G4 0NG, UK}

\author{Anthony Chefles}
\affiliation{Department of Physical Sciences, University of Hertfordshire, Hatfield, AL10 9AB, UK}

\author{Stephen M. Barnett}
\affiliation{Department of Physics and Applied Physics, University of Strathclyde, Glasgow, G4 0NG, UK}

\author{Erling Riis}
\affiliation{Department of Physics and Applied Physics, University of Strathclyde, Glasgow, G4 0NG, UK}

\author{Masahide Sasaki}
\affiliation{Communications Research Laboratory, Ministry of Posts and Telecommunications, Koganei, Tokyo 184-8795, Japan}

\date{March 1, 2001}

\begin{abstract}
We present the results of
generalized measurements of optical polarization designed to
provide one of three or four distinct outcomes.  This has allowed
us to discriminate between nonorthogonal polarization states with
an error probability that is close to the minimum allowed by
quantum theory.
Employing these optimal detection strategies on
sets of three (trine) or four (tetrad) equiprobable 
symmetric nonorthogonal polarization states,
we obtain a mutual information
that exceeds the maximum value attainable using conventional (von
Neumann) polarization measurements.
\end{abstract}



\maketitle

\section{Introduction}
\label{sec:intro}

The basic building block in quantum communications and information
is the qubit, that is a physical system with two orthogonal
quantum states.  A simple example is the horizontal and vertical
states of polarization associated with a single photon.  It is
also possible to prepare any superposition of vertical and
horizontal polarization and these correspond to other states of
linear, circular and elliptical polarization.  In quantum
communications a transmitting party (Alice) might select from a
number of possible nonorthogonal polarization states to send to
the receiving party (Bob).  This idea is the basis of the emerging
technology of quantum key distribution \cite{phoenix95a}.
Signal states with nonorthogonal polarizations may also arise as a
result of noisy communications channels acting on orthogonal signal
states. 
For certain types of noisy channels, the throughput of information
is actually maximized for nonorthogonal signal states \cite{fuchs97a}.

The message encoded in the transmitted photons must be retrieved
by measurement. This can be achieved either by a conventional (von
Neumann) measurement or by means of a generalized
measurement \cite{holevo73b,helstrom76a,peres93a}.
The von Neumann measurement gives
answers corresponding to one of a pair of orthogonal polarization
states.  The generalized measurements presented in this paper give
one of three or four possible outcomes corresponding to the
elements of a probability operator measure or POM \cite{helstrom76a},
also called a positive operator-valued measure \cite{peres93a}.

If there are only two possible states then it is
known that a single von Neumann measurement with two possible outcomes
will minimize the probability for error in identifying
the state \cite{helstrom76a,sasaki96a}.
State discrimination with this minimum allowed error probability has been
demonstrated for weak pulses of polarized light \cite{barnett97a}.
With only two possible states, it
is also possible to discriminate between the two possibilities without
error if we allow for the possibility of an inconclusive measurement
outcome \cite{ivanovic87a,dieks88a,peres88a,jaeger95a,chefles98a}.
Near error-free discrimination
between two nonorthogonal polarization states was first demonstrated
by Huttner \textit{et al}. \cite{huttner96a}.
More recently, we have shown that
it is possible to discriminate between nonorthogonal polarization states
\cite{clarke00a} with the probability for inconclusive outcomes at the
quantum limit determined by Ivanovic, Dieks and Peres.

\begin{figure}
    \begin{minipage}{\columnwidth}
    \begin{center}
        \resizebox{\columnwidth}{!}{\rotatebox{0}{\includegraphics{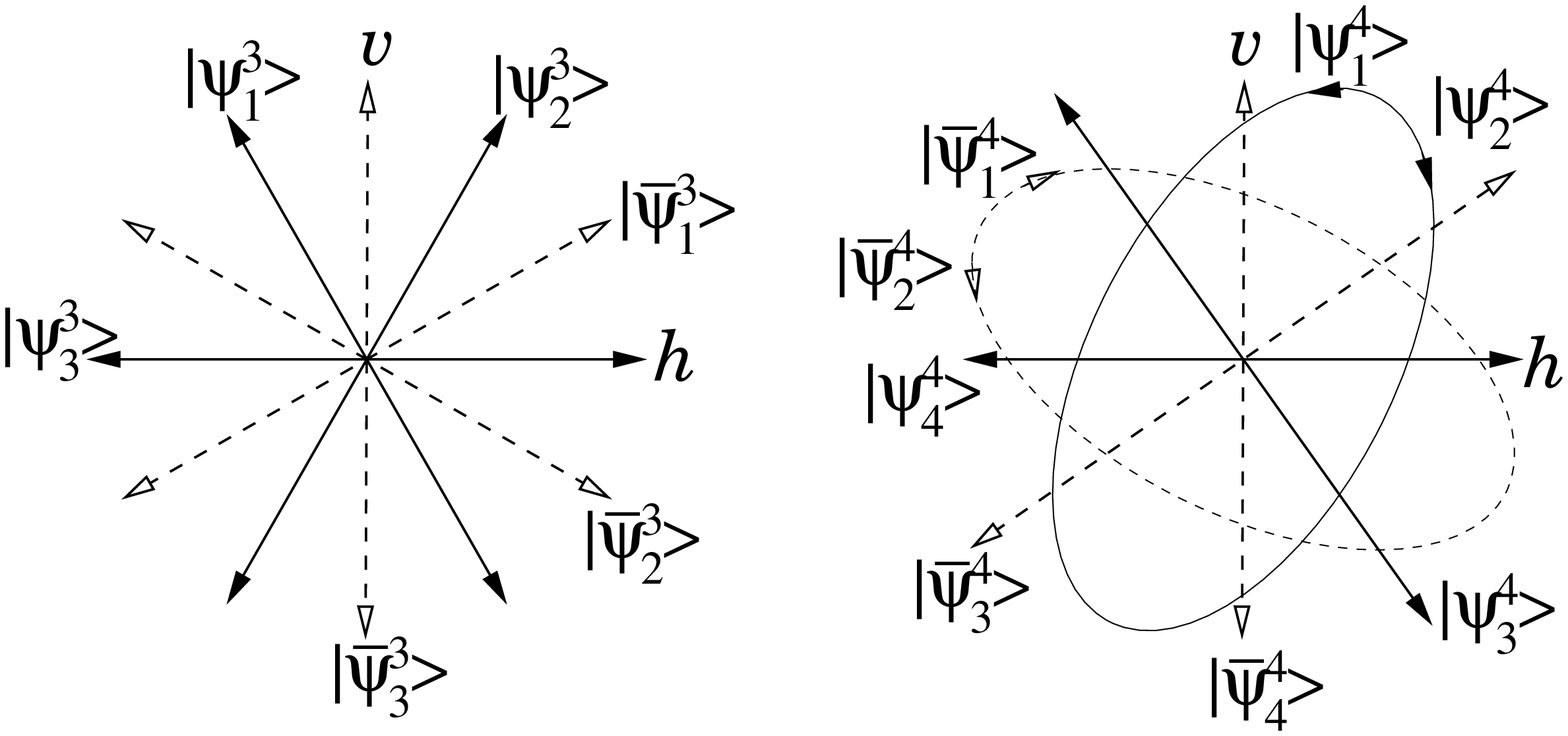}}}
    \end{center}
    \vspace{1em}
    \end{minipage}
    \hfill
    \begin{minipage}{\columnwidth}
    \begin{center}
        \resizebox{0.95\columnwidth}{!}{\rotatebox{0}{\includegraphics{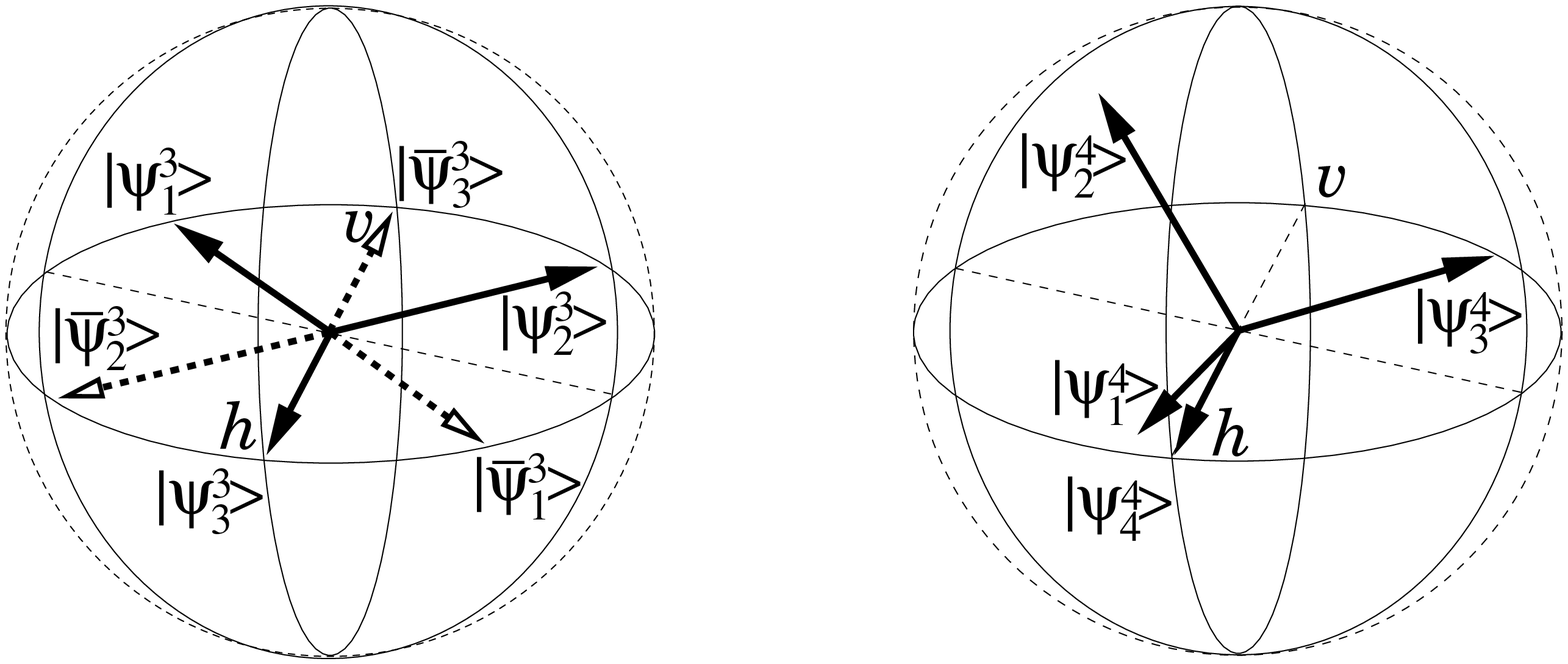}}}
    \end{center}
    \vspace{1em}
    \end{minipage}
    \hfill
    \begin{minipage}{\columnwidth}
    \begin{center}
        \resizebox{\columnwidth}{!}{\rotatebox{0}{\includegraphics{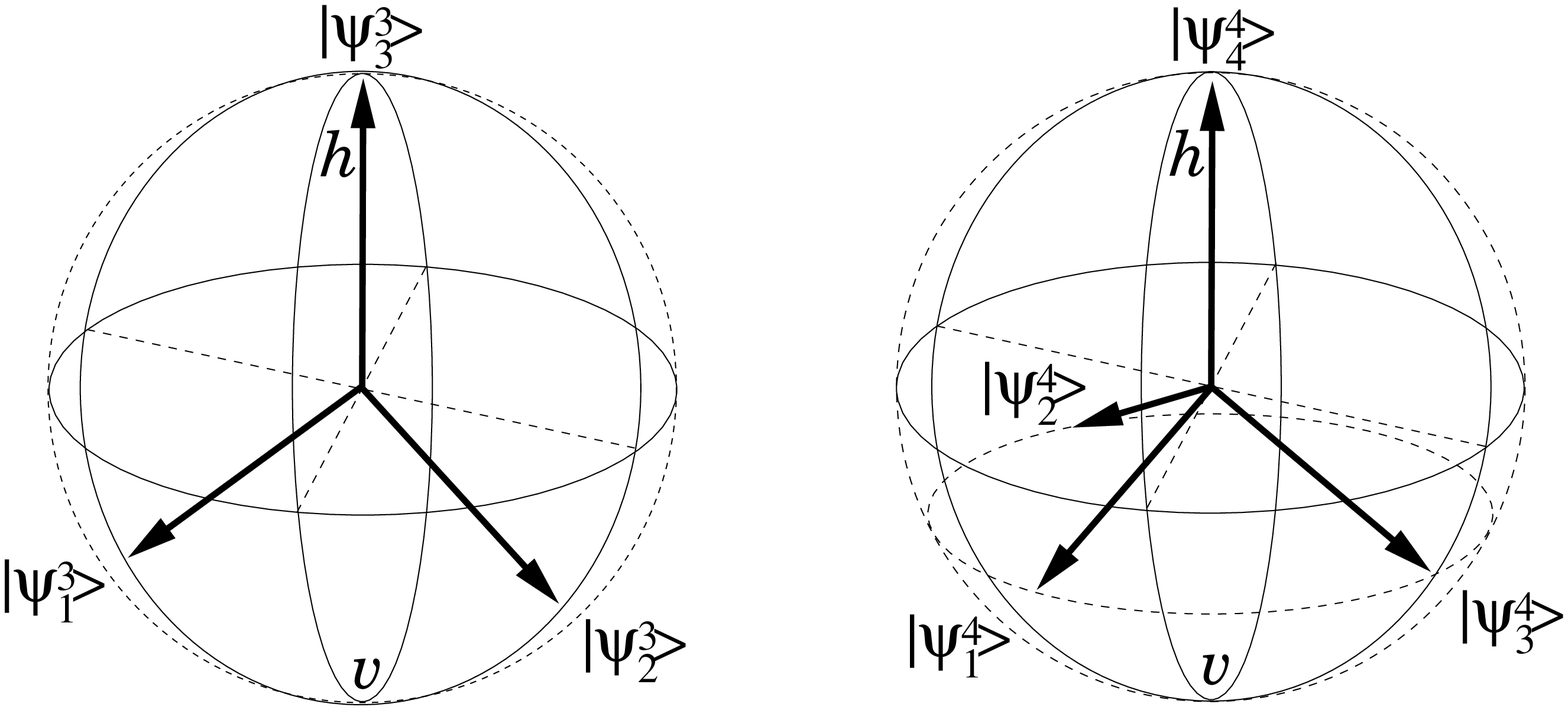}}}
    \end{center}
    \end{minipage}
    \caption{The trine (left) and tetrad (right) states shown as
         polarization directions (top), and on the
         Poincar\'{e} sphere (shown from two different orientations,
	 middle and lower).
         Anti-trine (orthogonal) states are also shown (dashed) in the top and
         middle left diagrams and antitetrad polarization states in 
         the top right diagram.}
    \label{fig:poincare}
\end{figure}

Unambiguous discrimination between two nonorthogonal states of a
qubit is an example of a generalized measurement \citep{helstrom76a,peres93a}
in that it requires three possible measurement outcomes but
there are only two orthogonal polarizations.  Measurements of
this kind are required if we wish to perform an optimum
measurement on a qubit prepared in more than two different states.
In this paper we will describe experiments on polarized light
prepared in one of three or four nonorthogonal states.  The three
states form a so-called trine ensemble \cite{holevo73b,peres92a,hausladen94a},
and take the form of three
states of linear polarization separated by an angle of $60^{\circ}$.
These states may be represented on the Poincar\'{e} sphere \cite{born87a}
as three equidistant points on the equator (see
Fig. \ref{fig:poincare}, left).  The four states form a tetrad
ensemble \citep{davies78a} and take the form of two linearly and
two elliptically polarized states. These states form the vertices
of a tetrahedron on the Poincar\'{e} sphere (see Fig.
\ref{fig:poincare}, right).

In this paper we describe our experimental realizations of two types of
optimal measurements on these trine and tetrad states.  The first gives
the minimum probability of error in identifying the
state \citep{helstrom76a,holevo73a,yuen75a,phoenix00a}.
The second provides the knowledge that the qubit was definitely
not prepared in one out of
the three or four possibilities, but does not discriminate further between
the remaining possibilities \cite{phoenix00a}.  This measurement is known to
give the accessible information (i.e., the mutual information maximized with 
respect to detection strategy) between transmitter and receiver for equal prior
probabilities for the trine states \cite{sasaki99a}.
We calculate the mutual information attainable from our experiments and show
that it exceeds that which can be obtained by the best von Neumann
measurement.

The rest of the paper is organized as follows.  In Section \ref{sec:theory}
the relevant theory of probability operator measures (POMs) is outlined.
In Section \ref{sec:optical} our implementation of such POMs
in optical networks is described theoretically, followed in Section
\ref{sec:expt} by presentation of the experimental setup and results.

\section{Theory}
\label{sec:theory}

Consider a communication channel in which photons are prepared,
with equal prior probabilities, in one of three or four known
states of polarization. Our task is to perform a measurement on
the light and so retrieve information about the prepared state. We
are interested in two types of measurement.  The first is designed
to minimize the probability of error in identifying the state
correctly, and the second is aimed at maximizing the mutual
information associated with the communications channel.

Our signal states are realizations of either the
trine \cite{holevo73b,peres92a,hausladen94a} or tetrad \cite{davies78a}
ensembles. The polarization states making up our trine
ensemble are
\begin{eqnarray}
|\psi_1^3\rangle &=& -\frac{1}{2}\left(|h\rangle
+\sqrt{3}|v\rangle\right), \nonumber \\
|\psi_2^3\rangle &=&
-\frac{1}{2}\left(|h\rangle -\sqrt{3}|v\rangle\right), \nonumber
\\ |\psi_3^3\rangle &=& |h\rangle,
\label{eq:trines}
\end{eqnarray}
corresponding to states of linear polarization separated by $60^{\circ}$
(see Fig. 1, left). The kets $|h\rangle$ and $|v\rangle$ are states of
horizontal and vertical polarization respectively.
The superscript label denotes the fact
that there are three members of the trine ensemble.  The
states making up our tetrad ensemble are
\begin{eqnarray}
|\psi_1^4\rangle &=& \frac{1}{\sqrt{3}}\left(-|h\rangle
        + \sqrt{2}\mathrm{e}^{-2\pi i/3}|v\rangle\right), \nonumber \\
|\psi_2^4\rangle &=& \frac{1}{\sqrt{3}}\left(-|h\rangle
        + \sqrt{2}\mathrm{e}^{+2\pi i/3}|v\rangle\right), \nonumber \\
|\psi_3^4\rangle &=& \frac{1}{\sqrt{3}}\left(-|h\rangle
        + \sqrt{2}|v\rangle\right), \nonumber \\
|\psi_4^4\rangle &=& |h\rangle,
\label{eq:tetrads}
\end{eqnarray}
where the superscript label again denotes the number of states in
the ensemble. The first two states correspond to elliptical
polarizations and the second two represent linear polarizations.
We also require the states making up the antitrine and antitetrad
ensembles.  These are the polarization states that are orthogonal
to the trine and tetrad states.  We denote the members of the
antitrine and antitetrad ensembles by an overbar.  The members of
the antitrine ensemble are
\begin{eqnarray}
|\bar{\psi}_1^3\rangle &=& \frac{1}{2}\left(\sqrt{3}|h\rangle - |v\rangle\right), \nonumber \\
|\bar{\psi}_2^3\rangle &=& -\frac{1}{2}\left(\sqrt{3}|h\rangle + |v\rangle\right), \nonumber
\\ |\bar{\psi}_3^3\rangle &=& |v\rangle,
\label{eq:antitrines}
\end{eqnarray}
and those of the antitetrad ensemble are
\begin{eqnarray}
|\bar{\psi}_1^4\rangle &=& -\frac{1}{\sqrt{3}}\left(
        \sqrt{2}\mathrm{e}^{2\pi i/3}|h\rangle
	+ |v\rangle \right), \nonumber \\
|\bar{\psi}_2^4\rangle &=& -\frac{1}{\sqrt{3}}\left(
        \sqrt{2}\mathrm{e}^{-2\pi i/3}|h\rangle
	+ |v\rangle \right), \nonumber \\
|\bar{\psi}_3^4\rangle &=& -\frac{1}{\sqrt{3}}\left(
        \sqrt{2}|h\rangle
	+ |v\rangle \right), \nonumber \\
|\bar{\psi}_4^4\rangle &=& |v\rangle.
\label{eq:antitetrads}
\end{eqnarray}

The optimum detection strategies require generalized measurements
which may be described in terms of the elements of a 
POM \cite{helstrom76a}.  Each of the possible results
($y_j$) of the measurement corresponds to a Hermitian operator, $\Pi_j$,
such that the probability of obtaining this result given that the
polarization was prepared in the state $|\psi^N_k\rangle$ is
\begin{equation}
P(y_j|\psi^N_k)=\langle \psi^N_k|\Pi_j|\psi^N_k\rangle,
\label{eq:conditional_prob}
\end{equation}
where $N$ is the number of states making up the ensemble (three for the
trine and four for the tetrad).

The requirement that the expectation value of $\Pi_j$ is a
probability leads to the constraints that all the eigenvalues of
$\Pi_j$ are positive or zero and that
\begin{equation}
\sum_j\Pi_j=\openone,
\label{eq:ident}
\end{equation}
where the sum runs over all possible results of the measurement.
This description of a measurement is more general than that
associated with a conventional (von Neumann) measurement.  For a
von Neumann measurement the POM elements are projectors onto the
orthonormal eigenstates of the measured observable.  In general,
however, the POM elements will be neither normalized nor
orthogonal.

\subsection{Minimum error probability}

The problem of finding the measurement which provides the minimum
error probability is an old one and it is instructive to present
it in a more general form than is strictly necessary for the
description of our experiments.  Consider a set of possible
prepared states with density matrices $\rho_k$ and let these occur
with prior probabilities $p_k$.  The lowest error probability will
be achieved by a measurement in which we associate a POM element
$\Pi_k$ with the statement that the prepared state was $\rho_k$.
The error probability is then
\begin{equation}
P_e=1-\sum_k p_k{\mathrm Tr}(\rho_k\Pi_k),
\label{eq:err_pr}
\end{equation}
where the sum gives the probability that the state will be 
correctly identified.
The necessary and sufficient conditions on the POM element for
this probability to attain its minimum value \cite{holevo73a,yuen75a} are
\begin{eqnarray}
\Pi_j(p_j\rho_j-p_k\rho_k)\Pi_k=\bm{0} \qquad &\forall&\,\,(j,k),
\label{eq:condition1} \\
\left(\sum_kp_k\rho_k\Pi_k\right)-p_j\rho_j \ge \bm{0} \qquad &\forall&\,\, j.
\label{eq:condition2}
\end{eqnarray}
The general solution for an arbitrary set of states and prior
probabilities is unknown but the solution is known for some
special cases.  We are interested in the pure states making up the
trine and tetrad ensembles.  These states are overcomplete in that
we can resolve the identity as a weighted sum of the
corresponding projectors
\begin{eqnarray}
\frac{2}{N}\sum_{k=1}^{N}|\psi_k^N\rangle\langle\psi_k^N|=\openone.
\label{eq:id_res}
\end{eqnarray}
If we impose the condition that the
states occur with equal prior probabilities then the POM elements
that minimize the error probability Eq. (\ref{eq:err_pr}) are \cite{yuen75a,ban97a}
\begin{eqnarray}
\Pi_k = \frac{2}{N}|\psi_k^N\rangle\langle\psi_k^N|.
\label{eq:pom_el}
\end{eqnarray}
This gives for the minimum error probability the value \footnote{
This treatment can be extended to higher dimensional
quantum systems.  The $N$ states form an overcomplete set in a state space of
$D$ 
dimensions if they satisfy Eq. (\ref{eq:id_res}) with the 2 replaced by $D$.  The
original 
POM elements are then given by Eq. (\ref{eq:pom_el}) with the 2 again replaced by $D$. 
This 
gives the minimum error probability $1 - D/N$ \cite{yuen75a}.
}
\begin{eqnarray}
P_e(\text{min}) = 1-\frac{2}{N}.
\end{eqnarray}
This is $\frac{2}{3}$ for the trine ensemble and $\frac{1}{2}$ for
the tetrad.
The remaining problem is to determine how we can
design experiments for which the POM elements are those given in
Eq. (\ref{eq:pom_el}).
Before addressing this problem we consider another scenario in
quantum signal detection.

\subsection{Mutual information}

The minimum error probability is used as a measure of how precisely
one can identify each signal states on average.
If instead of identifying each individual state as precisely as possible,
one is interested in extracting as much
classical information as possible from a sequence of quantum states,
a measure commonly used is the Shannon mutual information.

Let us suppose that Alice prepares the ensemble of quantum states
$\{|\psi^N_k\rangle\}$ to encode the classical message $\{x_k\}$,
where each element appears according to {\it a priori} probability
$p_k$. Let us call this ensemble $X=\{|\psi^N_k\rangle, p_k\}$.
The Shannon entropy of $X$ ,
\begin{equation}
H(X)=-\sum_k p_k \log p_k,
\label{H(X)}
\end{equation}
quantifies how much we know about $X$ just from the {\it a priori}
probabilities $p_k$.  Actually, the minimum value of $H(X)$ is zero,
corresponding to knowing everything about $X$, and the maximum
value of $H(X)$ corresponds to knowing nothing better than a random guess
for each element of $X$, so it is more natural to say that $H(X)$
measures what we {\it don't} know about $X$.
Bob now applies a measurement with $M$ possible outcomes $\{y_j\}$,
characterized by the POM elements $\{\Pi_j\}$.
The outcome, say, $y_j$, gives Bob more information about $X$.
The new probability distribution conditioned by $y_j$ is
\begin{equation}
P(\psi^N_k\vert y_j)=
   \frac{P(y_j\vert\psi^N_k)p_k}{P(y_j)},
\label{BayesRule}
\end{equation}
where $P(y_j)\equiv\sum_i P(y_j\vert\psi^N_i)p_i$ is
the probability of having $y_j$, and
$P(y_j\vert\psi^N_k)$ is defined by Eq. (\ref{eq:conditional_prob}).
One can then define the average conditional entropy by
\begin{equation}
H(X\vert Y)=-\sum_j P(y_j)
             \sum_k P(\psi^N_k\vert y_j) \log P(\psi^N_k\vert y_j).
\label{H(X|Y)}
\end{equation}
This quantifies the remaining uncertainty about $X$ after having
the knowledge of the conditioning variable $Y=\{y_j, P(y_j)\}$.
The information gained as a result of making the measurement $Y$
is naturally defined by
\begin{eqnarray}
I(X:Y)&=&H(X)-H(X\vert Y), \\
      &=&\sum_{j=1}^M \sum_{k=1}^N
        \left[ \rule[-1.3em]{0em}{2.5em} \right.
        p_k P(y_j\vert \psi^N_k) \nonumber \\*
       \times&& \left.\log_2 \left( \frac{P(y_j\vert \psi^N_k)}
        {\sum_{k'=1}^N p_{k'} P(y_j\vert \psi^N_{k'})} \right)
        \right].
\label{I(X:Y)eq}
\end{eqnarray}
This is the Shannon mutual information between $X$ and $Y$.
In general, minimizing the average error probability and
maximizing the mutual information are different problems,
and in fact, one often finds the optimal POMs are different in each case.
To extract as much information as possible, Bob has to maximize
the mutual information with respect to the POM.
The maximum value
\begin{equation}
I_{\rm Acc}=\max_{\{\Pi_j\}}I(X:Y),
\end{equation}
is called the {\it accessible information} of the ensemble $X$.
To obtain the accessible information, it is not necessary for the
number of measurement outcomes to be the same as the number of signal states.
(In the experiments described in this paper, however,
these will always be equal.)

The trine and tetrad ensembles comprise nonorthogonal states and
hence errors in determining each individual state are inevitable.
It is possible, however, to achieve asymptotically error-free
transmission of information by employing block coding schemes
based on sequences of the states.
The accessible information has a practical meaning in this problem.
Let $\{\Pi_j\}$ be the POM attaining the accessible information for
$X$.
We consider block sequences of $\{|\psi^N_k\rangle\}$ of length $n$
and choose $2^k$ such sequences (codewords) to code classical messages.
Suppose that Bob applies $\{\Pi_j\}$ on each states of a received
block sequence separately, and guess the codeword sent based on the
outcome with any classical procedure.
Then if $k/n<I_{\rm Acc}$, Alice and Bob can
communicate with an arbitrarily small error by choosing $n$ large
enough \cite{cover91a}.

Thus it is important to find the optimal POM and,
in particular, to demonstrate its physical implementation
in practice, for developing communication technology though a
quantum limited channel.
The general problem of finding the accessible information for any
given set of signal states remains unsolved.
The complete results are known for a few special cases
\cite{davies78a,sasaki99a,osaki00a}.
For the trine states with equal prior
probabilities the accessible information is attained for a
measurement with POM elements based on the antitrine states
\begin{eqnarray}
\Pi_j = \frac{2}{3}|\bar{\psi}_j^3\rangle\langle\bar{\psi}_j^3|.
\end{eqnarray}

The optimality of this measurement strategy was conjectured in
\cite{holevo73a,peres92a} and finally proven by
Sasaki \textit{et al}. \cite{sasaki99a}.
The corresponding accessible information is $\log_2
\frac{3}{2}=0.585$ bits.  This clearly exceeds the value
$(-\frac{1}{3} + \frac{1}{2} \log_2 3)=0.459$ bits which is the
maximum mutual information attainable for a von Neumann
measurement corresponding to finding one of two orthogonal
polarizations.

The accessible information for the tetrad ensemble is not so strongly
established as it is for the trine ensemble.  Davies \cite{davies78a}
has conjectured that it corresponds to the POM elements
\begin{eqnarray}
\Pi_j = \frac{1}{2}|\bar{\psi}_j^4\rangle\langle\bar{\psi}_j^4|
\end{eqnarray}
associated with the antitetrad states.  The mutual information
associated with this measurement strategy (conjectured to be the
accessible information) is $\log_2 \frac{4}{3}=0.415$ bits.  This
again exceeds the value of $\frac{3}{2}(1-\frac{1}{2} \log_2
3)=0.311$ bits attained for the best possible von Neumann
measurement.

It should be noted that the measurement strategy based on the antitrine
and antitetrad states only identifies the correct state with
probability $1/(N-1)$.  However, it obtains more information than the minimum error strategy by identifying for certain
one of the states that was {\it not} sent\footnote{
In the antitrine case, this measurement strategy is related to
the optimal discrimination between two nonorthogonal states
\cite{ivanovic87a,dieks88a,peres88a,jaeger95a,chefles98a}, 
demonstrated experimentally in \cite{huttner96a,clarke00a}.
If the state to be identified is known to be one of only two of the
three possible antitrine states, either $|\bar{\psi}^3_j\rangle$
or $|\bar{\psi}^3_k\rangle$, then the three possible outcomes
correspond to ``not-$|\bar{\psi}^3_j\rangle$''
(i.e. ``is $|\bar{\psi}^3_k\rangle$''), ``not-$|\bar{\psi}^3_k\rangle$'',
and ``inconclusive''.
}.

\section{Optical implementation}
\label{sec:optical}

Having described the results that are predicted for various optimal
measurements of the trine and tetrad states, we now turn to
the implementation of such optimal generalized measurements
using specific optical networks to measure photon polarization
states.  The theoretical operation of the optical networks will
be described in the next subsections, followed in Section
\ref{sec:expt} by the description and results of the
actual experiments we performed.

The generalized measurements described in the previous section have to be
implemented in practice as von Neumann measurements with
simple yes/no results, but in an enlarged state space
\cite{naimark40a,holevo73a,helstrom76a,peres93a}.
In our experiments, the state space is enlarged by incorporating
an interferometer into the measurement apparatus.
The enlargement of the state space is
realized by the introduction of vacuum modes entering through the unused 
input port.
A single interferometer allows up to
four mutually exclusive (orthogonal) possible results from a single
photon input state.

In order to describe clearly how our apparatus works, we first set out our
sign conventions for the optical components.
The main polarizing beam splitters
are oriented so that horizontally polarized photons, $|h\rangle$,
go straight through, while vertically polarized photons, $|v\rangle$,
are deflected through an angle of $\pi/2$.
For the non-polarizing beam splitter, the transmission and reflection
coefficients are assumed to be equal (both $1/\sqrt{2}$), and since
there is only one (non-vacuum) input to it, we take there to be
no phase difference between the two outputs.
For the quarter and half waveplates, their effect when placed with their
fast axis at an arbitrary angle (measured anticlockwise from
the horizontal viewed in the direction of travel of the photons)
can be described in terms of Jones matrices (see Appendix \ref{app:jones}).
In particular, a half waveplate placed with its
axes at an angle of $\pi/4$ to the vertical/horizontal
rotates $|v\rangle$ to $|h\rangle$, and $|h\rangle$ to $|v\rangle$,
as is well known.
A half waveplate with axes at $\pi/8$ to the vertical/horizontal
and a quarter waveplate with axes
at an angle of $\pi/4$ both produce maximum mixing between
the $|v\rangle$ and $|h\rangle$ states, which is what is required
to analyze the state obtained when the two arms of the interferometer
are recombined.  The quarter waveplate is used in the tetrad case,
where there is a phase factor of $i$ between the components in each arm.
A half waveplate placed
at the special angle $\alpha/2 = \frac{1}{2}\arcsin(1/\sqrt{3})\simeq
17.63^{\circ}$, is used to rotate the photon until the amplitude of the
$|h\rangle$ component is reduced to $\sqrt{2/3}$ of its initial value.

\subsection{Trine}

\begin{figure}
    \begin{minipage}{\columnwidth}
    \begin{center}
        \resizebox{0.8\columnwidth}{!}{\includegraphics{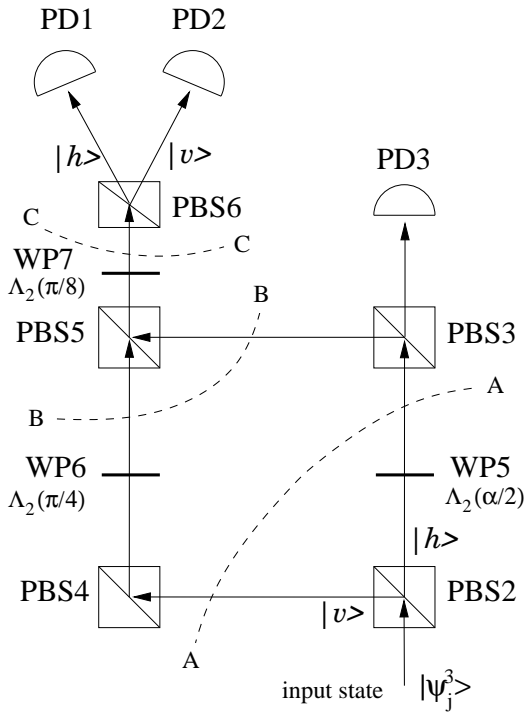}}
    \end{center}
    \end{minipage}
    \caption{Optical network for the trine generalized  measurements.
    PBS = polarizing beam splitter, WP = waveplate
    and PD = photodetector
    (labels correspond with Fig. \protect{\ref{fig:expt}}).
    The dotted lines AA, BB and CC are
    referred to in the text.  The Jones matrix corresponding to each
    waveplate is shown alongside it ($\Lambda_2(\alpha/2)$, etc.)
    with the angle indicating the orientation of the axes.}
    \label{fig:trine}
\end{figure}

An optical network
that realizes the minimum error probability for the three trine states is
depicted in Fig. \ref{fig:trine}.
This is a variation of recently proposed networks \cite{sasaki99a,phoenix00a}.
Our trine input states are represented in terms of photon polarization states
as in Eq. (\ref{eq:trines}),
\begin{eqnarray}
|\psi^3_1\rangle &=& -\frac{1}{2}\left(|h\rangle +\sqrt{3}|v\rangle\right), \nonumber \\
|\psi^3_2\rangle &=& -\frac{1}{2}\left(|h\rangle -\sqrt{3}|v\rangle\right), \nonumber \\
|\psi^3_3\rangle &=& |h\rangle. \nonumber
\end{eqnarray}
After passing through the polarizing beam splitter PBS1, the $|h\rangle$
component of each state goes into the upper arm of the interferometer
and the $|v\rangle$ component goes into the lower arm.  The
half-waveplate WP5 then rotates the $|h\rangle$ component such that
at the stage specified by the line AA in Fig. \ref{fig:trine}, the states
have been transformed to
\begin{eqnarray}
|\psi^3_1\rangle_{\mathrm A} &=& -\frac{1}{2}\frac{1}{\sqrt{3}}
    \left(\rule[-0.2em]{0em}{1.2em}     
    \sqrt{2}|h_{\mathrm U}\rangle + |v_{\mathrm U}\rangle\right)
    -\frac{\sqrt{3}}{2}|v_{\mathrm L}\rangle, \nonumber \\
|\psi^3_2\rangle_{\mathrm A} &=& -\frac{1}{2}\frac{1}{\sqrt{3}}
    \left(\rule[-0.2em]{0em}{1.2em}     
    \sqrt{2}|h_{\mathrm U}\rangle + |v_{\mathrm U}\rangle\right)
    +\frac{\sqrt{3}}{2}|v_{\mathrm L}\rangle, \nonumber \\
|\psi^3_3\rangle_{\mathrm A} &=& \frac{1}{\sqrt{3}}
    \left(\rule[-0.2em]{0em}{1.2em}         
    \sqrt{2}|h_{\mathrm U}\rangle + |v_{\mathrm U}\rangle\right),
\end{eqnarray}
where the subscripts U and L denote the upper and lower arms of the
interferometer respectively.
The $|h_{\mathrm U}\rangle$ part of these states passes out of the
interferometer through PBS2 and into photodetector PD3.
If the input state was $|\psi^3_3\rangle$, there is a
probability of $\frac{2}{3}$ that PD3 will detect the photon.
If the input state was $|\psi^3_1\rangle$ or $|\psi^3_2\rangle$,
there is a probability
of $\frac{1}{6}$ for each state that PD3 will detect the photon.
Turning this round, if PD3 detects the photon, there is
a probability of $\frac{2}{3}$ that the input state was
$|\psi^3_3\rangle$ and a probability of $\frac{1}{6}$ each that
the input state was $|\psi^3_1\rangle$ or $|\psi^3_2\rangle$.

Continuing through the interferometer, the half waveplate WP6
rotates the component in the lower arm 
from $|v_{\mathrm L}\rangle$ to $|h_{\mathrm L}\rangle$, so at the stage
specified by BB in Fig. \ref{fig:trine}, the states are
\begin{eqnarray}
|\psi^3_1\rangle_{\mathrm B} &=& -\frac{1}{\sqrt{6}}|PD3\rangle
    -\frac{1}{2}\left(\frac{1}{\sqrt{3}}|v_{\mathrm U}\rangle
        + \sqrt{3}|h_{\mathrm L}\rangle\right), \nonumber \\
|\psi^3_2\rangle_{\mathrm B} &=& -\frac{1}{\sqrt{6}}|PD3\rangle
    -\frac{1}{2}\left(\frac{1}{\sqrt{3}}|v_{\mathrm U}\rangle
        - \sqrt{3}|h_{\mathrm L}\rangle\right), \nonumber \\
|\psi^3_3\rangle_{\mathrm B} &=& \frac{\sqrt{2}}{\sqrt{3}}|PD3\rangle
    + \frac{1}{\sqrt{3}}|v_{\mathrm U}\rangle.
\end{eqnarray}
It is necessary to arrange the path lengths in the arms of the interferometer
such that there is a relative phase of $\pi/2$ between the U and L
components.  Then, after the polarizing beam splitter PBS4 recombines
the two components, the half waveplate WP9 mixes the
two polarizations such that, at the stage indicated by CC on
Fig. \ref{fig:trine}, the states become
\begin{eqnarray}
|\psi^3_1\rangle_{\mathrm C} &=& -\frac{1}{\sqrt{6}}|PD3\rangle
    - \frac{1}{\sqrt{3}}\left(\sqrt{2}|h\rangle
    + \frac{1}{\sqrt{2}}|v\rangle\right), \nonumber \\
|\psi^3_2\rangle_{\mathrm C} &=& -\frac{1}{\sqrt{6}}|PD3\rangle
    + \frac{1}{\sqrt{3}}\left(\frac{1}{\sqrt{2}}|h\rangle
    + \sqrt{2}|v\rangle\right), \nonumber \\
|\psi^3_3\rangle_{\mathrm C} &=& \frac{\sqrt{2}}{\sqrt{3}}|PD3\rangle
    + \frac{1}{\sqrt{6}}\left(|h\rangle - |v\rangle\right),
\end{eqnarray}
Beam splitter PBS6 separates the $|h\rangle$ and $|v\rangle$
components so they go into detectors PD1 and PD2 respectively.
From this we can see that if PD1 detects
the photon, there is a probability of $\frac{2}{3}$ that the input state was
$|\psi^3_1\rangle$, and a probability of $\frac{1}{6}$ each that it was
$|\psi^3_2\rangle$ or $|\psi^3_3\rangle$,
similarly for PD2 with states $|\psi^3_1\rangle$ and $|\psi^3_2\rangle$
interchanged.
In other words, after passing through this optical network, the final
states (denoted by subscript F) are given by
\begin{eqnarray}
|\psi^3_1\rangle_{\mathrm F} &=& -\frac{1}{\sqrt{6}}|PD3\rangle
    - \frac{\sqrt{2}}{\sqrt{3}}|PD1\rangle
    - \frac{1}{\sqrt{6}}|PD2\rangle), \nonumber \\
|\psi^3_2\rangle_{\mathrm F} &=& -\frac{1}{\sqrt{6}}|PD3\rangle
    + \frac{1}{\sqrt{6}}|PD1\rangle
    + \frac{\sqrt{2}}{\sqrt{3}}|PD2\rangle, \nonumber \\
|\psi^3_3\rangle_{\mathrm F} &=& \frac{\sqrt{2}}{\sqrt{3}}|PD3\rangle
    + \frac{1}{\sqrt{6}}|PD1\rangle
    - \frac{1}{\sqrt{6}}|PD2\rangle.
\end{eqnarray}
This is all normalized so the probabilities are given by the
square of the amplitudes of each photodetector state.
(The photodetector states $|PD1\rangle$, $|PD2\rangle$ and $|PD3\rangle$
are necessarily mutually orthogonal as only one photodetector out of the
three can detect the photon.)

In order to implement the POM corresponding to the maximum mutual information,
instead of constructing the optical network that corresponds to 
POM elements $|\bar{\psi}^3_j\rangle\langle \bar{\psi}^3_j|$,
and applying it to the trine states, we use the optical network
just described (corresponding to the POM elements $|\psi^3_j\rangle\langle
\psi^3_j|$) and apply it to the antitrine states $|\bar{\psi}^3_j\rangle$.
This is obviously completely equivalent theoretically and more
practical experimentally.
We find that the state $|\bar{\psi}^3_j\rangle$ should never trigger
photodetector $j$.
It will, however, lead to detection in either of the remaining two
photodetectors with equal probability ($\frac{1}{2}$) \cite{phoenix00a}.

\subsection{Tetrad}

\begin{figure}
    \begin{minipage}{\columnwidth}
    \begin{center}
        \resizebox{0.9\columnwidth}{!}{\includegraphics{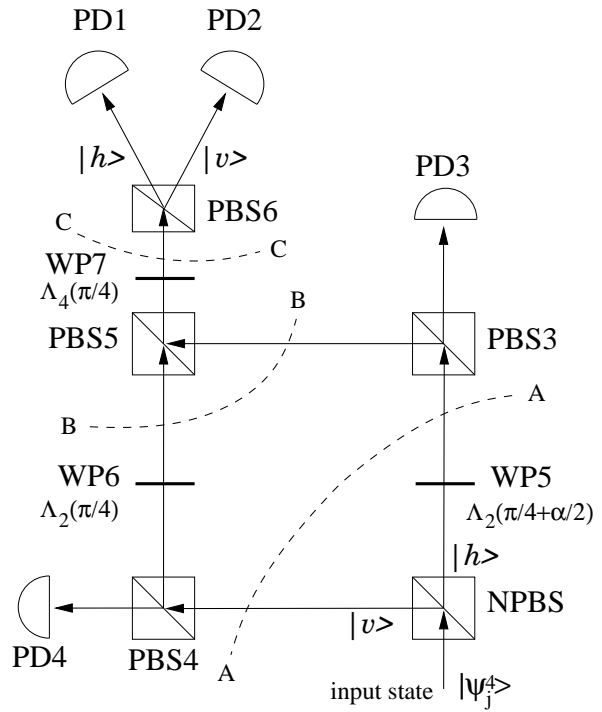}}
    \end{center}
    \end{minipage}
    \hfill
    \caption{Optical network for the tetrad generalized measurement.
    NPBS = non-polarizing beam splitter,
    PBS = polarizing beam splitter, WP = waveplate 
    and PD = photodetector
    (labels correspond with Fig. \protect{\ref{fig:expt}}).
    The dotted lines AA, BB and CC are
    referred to in the text.  The Jones matrix corresponding to each
    waveplate is shown alongside it ($\Lambda_2(\alpha/2)$, etc.)
    with the angle indicating the orientation of the axes.}
    \label{fig:tetrad}
\end{figure}

The optical network we used to implements the optimal measurement for the
tetrad states is shown in Fig. \ref{fig:tetrad}.
It is similar to the trine network in Fig. \ref{fig:trine}, with an extra
detector PD4.  The key differences are a non-polarizing beam splitter (NPBS) in 
place of PBS2 and WP5 rotated to an angle of $(\pi/2 +\alpha)/2$.
The four tetrad states are represented as polarized photons as in Eq.
(\ref{eq:tetrads}),
\begin{eqnarray}
|\psi^4_1\rangle &=& \frac{1}{\sqrt{3}}\left(-|h\rangle
        + \sqrt{2}\mathrm{e}^{-2\pi i/3}|v\rangle\right), \nonumber \\
|\psi^4_2\rangle &=& \frac{1}{\sqrt{3}}\left(-|h\rangle
        + \sqrt{2}\mathrm{e}^{+2\pi i/3}|v\rangle\right), \nonumber \\
|\psi^4_3\rangle &=& \frac{1}{\sqrt{3}}\left(-|h\rangle
        + \sqrt{2}|v\rangle\right), \nonumber \\
|\psi^4_4\rangle &=& |h\rangle. \nonumber
\end{eqnarray}

The tetrad network can be understood most straightforwardly by noting the effect
of the half waveplate, WP5, on each of the four tetrad states.
Denoting the operation of WP5 by the Jones matrix $\Lambda_2(\pi/4+\alpha/2)$,
\begin{eqnarray}
\Lambda_2(\pi/4+\alpha/2)|\psi^4_1\rangle &=& i|\psi^4_2\rangle, \nonumber \\
\Lambda_2(\pi/4+\alpha/2)|\psi^4_2\rangle &=& -i|\psi^4_1\rangle, \nonumber \\
\Lambda_2(\pi/4+\alpha/2)|\psi^4_3\rangle &=& |\psi^4_4\rangle, \nonumber \\
\Lambda_2(\pi/4+\alpha/2)|\psi^4_4\rangle &=& |\psi^4_3\rangle,
\end{eqnarray}
so that, apart from phase factors, state $|\psi^4_1\rangle$ is converted
into $|\psi^4_2\rangle$ and vice versa, similarly for states
$|\psi^4_3\rangle$ and $|\psi^4_4\rangle$.
Remembering the first beam splitter is $50/50$ non-polarizing,
the network is thus essentially symmetrical in its operation.
The full details are given in Appendix \ref{app:tetrads}.
The final result is that state $|\psi^4_1\rangle$ reaches
photodetector PD1 with probability $\frac{1}{2}$ and each of the remaining 
photodetectors with probability $\frac{1}{6}$.
Similarly each of the remaining three states
will trigger its associated photodetector with probability $\frac{1}{2}$
and the others with probability $\frac{1}{6}$.

The maximum mutual information measurement is realized in the same
way as for the trine ensemble, by using the antitetrad states as input.
If the antitetrad states are introduced into this network then we find 
that the state $|\bar{\psi}^4_j\rangle$ should never trigger photodetector $j$.
It will, however, lead to detection in any of the remaining
three photodetectors with equal probability ($\frac{1}{3}$).

\section{Experiments}
\label{sec:expt}

\begin{figure}
    \begin{minipage}{\columnwidth}
    \begin{center}
        \resizebox{\columnwidth}{!}{\includegraphics{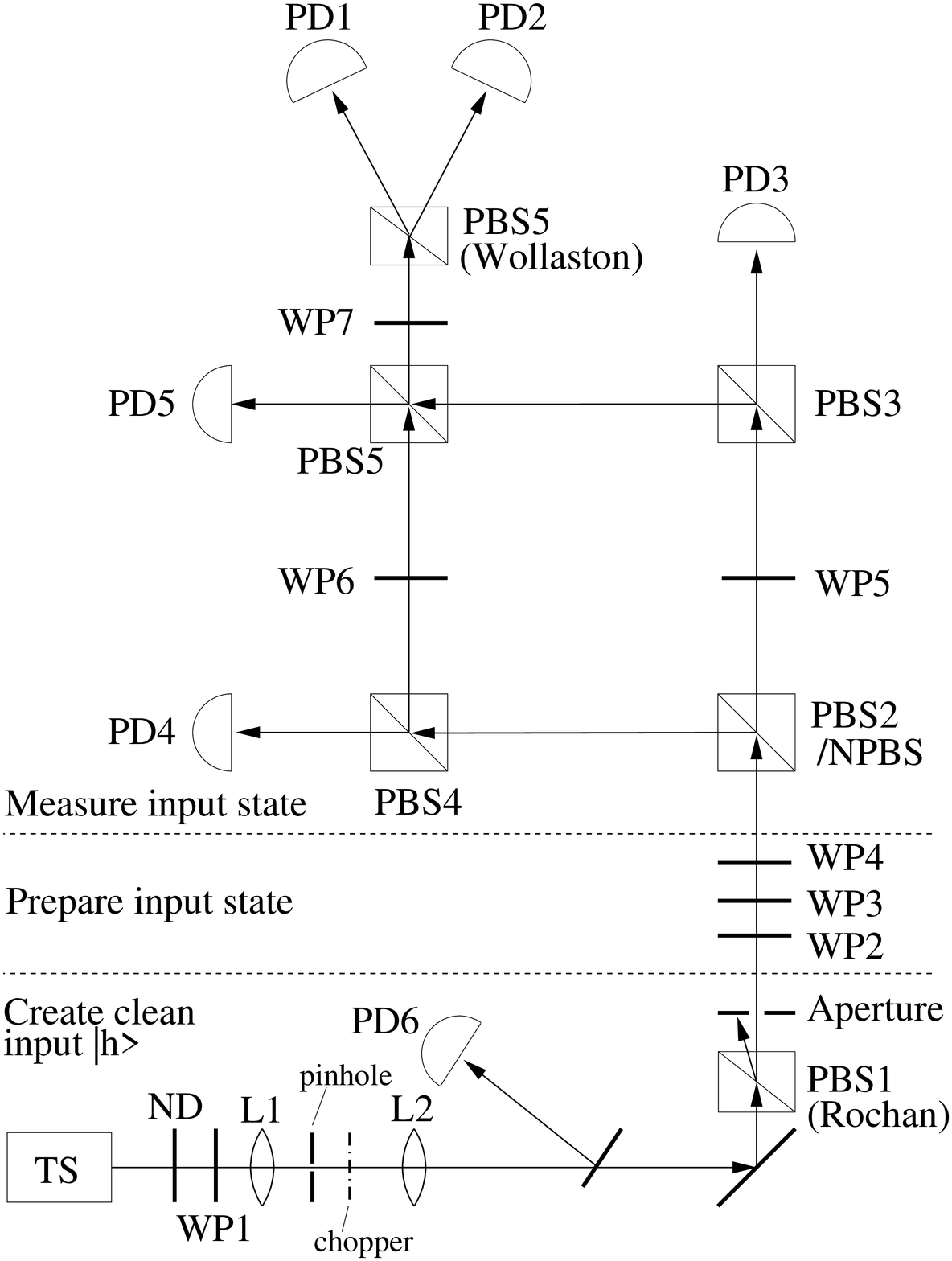}}
    \end{center}
    \end{minipage}
    \hfill
    \caption{Experimental setup for trine and tetrad state measurements.
    See text for full description.  L = lens, ND =
    neutral density filter to attenuate the light, NPBS = non-polarizing
    beamsplitter, PBS = polarizing beam splitter, PD = photodetector,
    TS = Titanium-Sapphire laser, WP = waveplate, either
    half or quarter.}
    \label{fig:expt}
\end{figure}

Figure \ref{fig:expt} shows the experimental arrangements used for
the trine and tetrad experiments.  The apparatus was easily
converted between the two experiments by changing WP7 from a half to a quarter
waveplate and exchanging PBS2 with NPBS.
We first discuss the general arrangement and
techniques that are common to both experiments (see also Ref.
\cite{clarke00a}, where the same experimental setup is described in
more detail).  The individual
details and results of the two experiments are then described
separately in Sections \ref{sec:trine_expt} and \ref{sec:tetrad_expt}.

The light source was a mode-locked light Ti:Sapphire laser
operating at 780 nm with a repetition rate of 80.3 MHz.  The pulse
duration of 1 ps corresponds to a pulse length of 300 $\mu$m. The
repetition rate of the laser ensured that there was only one pulse
in the optical system at any one time.  The length of each pulse
was much shorter than the path length of the interferometer. This
meant that each pulse could only interact with one optical component
at a time.

The laser light was passed through a 60 $\mu$m pinhole to produce
a spherical wavefront.  Lens L2 then focused the light onto the
photodiodes PD1-5 (Centronix, BPX65) through the interferometer.
This arrangement minimized the phase distortions across the
wavefront as it passed through the non-ideal interfaces of the
beamsplitters, maximizing the visibility of the interference. The focus on
the detectors was much smaller than the 1 mm$^2$ detector area,
yielding almost 100\% spatial collection of the light.
Losses due to the anti-reflective coatings on the
beamsplitters were also small, so almost all the photons
entering the interferometer reached the detectors.

The Rochon polarizing beamsplitter PBS1 was used to polarize the
light with a linearity of better than 5000:1.  The Wollaston
polarizing beamsplitter PBS6 was chosen for its similar efficient
polarizing properties. The waveplates WP2-4 were then used to
prepare the input states, see Appendix
\ref{app:jones}.

In these experiments, we measure the average current from the
photodiodes rather than counting discrete photon events.  To
detect the small quantities of light, phase sensitive detection
\cite{horowitz89a} was employed using a chopper wheel, differential
amplifier and a lock-in amplifier. The differential amplifier was used
to reduce the noise of the signal by canceling out common ground loop noise
using a darkened detector. The detectors PD1-5 had a nominal
quantum efficiency of 83\% at 780 nm and were terminated by 10 M$\Omega$. At
an average laser intensity of 0.1 photons per pulse this
corresponds to a detected voltage of 11 $\mu$V.  With time
constants of 10 to 30 seconds, light levels of 0.01 photons per
pulse were detectable with an accuracy of approximately 1\% and an
uncertainty of approximately $\pm$2.5\%.

Neutral density filters and rotation of the waveplate WP1, in
conjunction with PBS1, attenuated the light entering the
interferometer to an average of 0.1 photons per pulse. A pick-off
beam was measured on PD6 using phase sensitive detection with a
separate lock-in amplifier. This photodiode was used to normalize
output amplitude variations of the laser light when monitoring PD1-5
during all measurements and calibrations. These detectors,
supplied in parallel by a single 9 V source, were calibrated
relative to each other to better than 1\% by changing the
distribution of light around the interferometer.

All the waveplates used in the experiments were housed in
specially designed mounts.  The waveplate was freely rotated to
zero its position, measured optically, and locked in position.  It
could subsequently be moved by predefined fixed angles, including
$17.63^{\circ}$ and $27.37^{\circ}$, with an accuracy of $\pm0.05^{\circ}$ and
repeatability of $\pm 0.01^{\circ}$.  This enabled quick and
accurate changes to the input polarization states in the trine and
tetrad experiments. The waveplates were measured to maintain the
linearity of the polarization to 1 part in 2000.

The interferometer for the trine experiment was identical to that
described in \cite{clarke00a}.  It was constructed from four
polarizing beamsplitters mounted on a machined monolithic
aluminum block.  In a conventional Mach-Zehnder operation, the
extinction ratio of the output of the interferometer on detectors
PD1 and PD2 was regularly measured to be 200:1.  The system was
stable enough to be left for over half an hour without
significantly impairing this extinction ratio.

For the tetrad experiment the first beamsplitter, NPBS, was
non-polarizing with a nominally equal splitting ratio at 780 nm.
As the splitting ratio was highly wavelength dependent, the
Ti:Sapphire laser was tuned to obtain the most equal splitting
ratio, which was estimated to be $49.5:50.5$.

\subsection{Trine Experiment}
\label{sec:trine_expt}

The interferometer was aligned by using it as a conventional Mach-Zehnder
interferometer, with high light powers ($>>1$ photon per pulse).
The input polarization was set to be linear at $45^{\circ}$ to the
horizontal and WP5 was rotated to produce vertically polarized light. 
After alignment, the light entering the interferometer was attenuated
to 0.1 photons per pulse and WP5 was rotated by $17.63^{\circ}$
so that the polarization of the horizontal component transmitted
by PBS2 was rotated anti-clockwise $35.27^{\circ}$ from the
horizontal.  This reduced the amplitude of the horizontal component
to $\sqrt{2/3}$ of its initial value. The 6 input trine and
antitrine states were constructed by rotating waveplate WP2 in 15
degree intervals (WP3,4 are not present in this experiment). The
results were obtained by taking measurements for one detector
(PD1-3) as the 6 input states were changed.  Then the next
detector was measured for the same 6 states.

\begin{figure}
    \begin{minipage}{\columnwidth}
        \resizebox{\columnwidth}{!}{\includegraphics{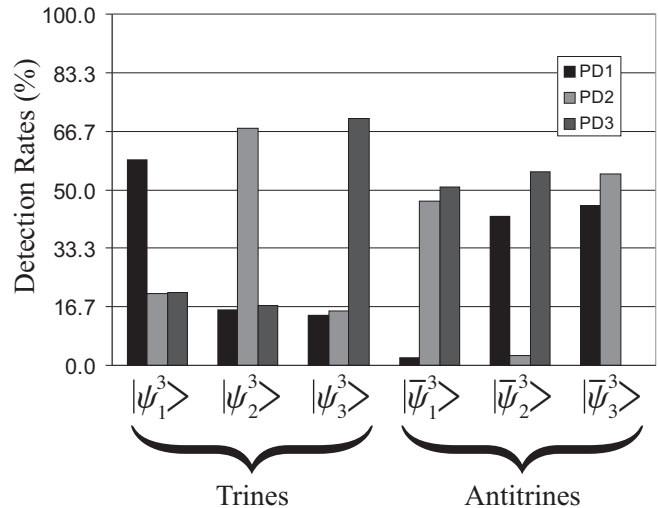}}
    \end{minipage}
    \hfill
    \caption{Histogram showing the trine and antitrine results.
    Normalization ensures the sum of counts in detectors 1, 2 and 3 is $100\%$
    for each input state.
    The theoretical ratios are
    $\frac{2}{3}:\frac{1}{6}:\frac{1}{6}$ for the trine and
    $\frac{1}{2}:\frac{1}{2}:0$
    for the antitrine states.}
    \label{fig:trine-results}
\end{figure}

After the relative detector calibrations were taken into account,
we normalized the results for each input state such that the total
average measured probability on photodiodes PD1-3 summed to unity.
The $\approx$1\% total leakage from PBS3 and PBS4 into detectors
PD4 and PD5 is not significant.  For the purposes of mutual
information it is the ratio of counts in photodiodes PD1-3 that is
important, because it is these detectors that
distinguish between the input states. The results are shown in Fig.
\ref{fig:trine-results}.

\begin{figure}
    \begin{minipage}{\columnwidth}
        \resizebox{\columnwidth}{!}{\includegraphics{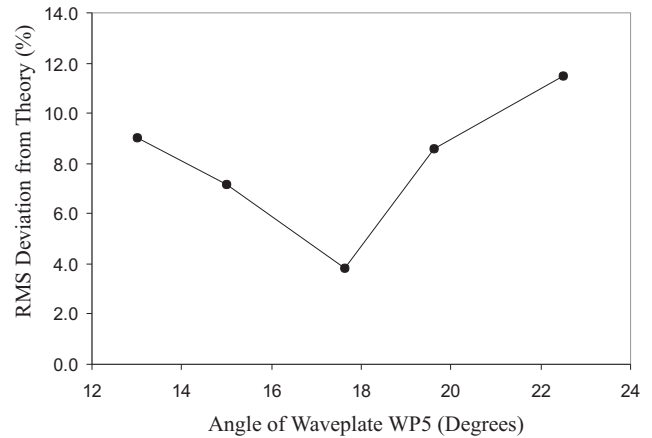}}
    \end{minipage}
    \hfill
    \caption{The angle of WP5 is varied around the theoretical value
    of $\alpha/2=17.63^{\circ}$.  The graph shows the RMS deviation
    of the results from the optimum theoretical results.}
    \label{fig:rmsdev}
\end{figure}

The dependence of the measurements on the angle of WP5 was then
investigated.  The experiment was repeated as the angle of the WP5
was varied either side of the theoretical value of $17.63^{\circ}$.
The alignment of the interferometer was checked and adjusted, if
necessary, before taking data at each separate angle.  Figure
\ref{fig:rmsdev} shows the root mean squared deviation of the 18
trine and antitrine measurements from the ideal values for each
angle.

Before we discuss the results it is useful to review the sources
of error.  The largest is the noise of the detector signal and
from the phase sensitive detection processing.  The observed
uncertainty in the measurements of the output ports where a zero
count rate is expected in the antitrine experiment is
approximately $2.5\%$. This level of error is often similar to
the average signal present.  We therefore conclude that the zero
count measurements are limited by the detector noise if the
measured signal is of order 2\%. This has important implications
for the errors in the derived mutual information, see Section
\ref{sec:mutual-info}. The noise is mainly attributable to the
weakness of these signals in comparison to the magnitude of ground
loop and stray light noise, and also variations in the chopping
wheel frequency leading to varying offsets.  We estimate that the
error due to the amplitude normalization procedure involving PD6
is less than 0.5\%, which is much smaller than the Ti:Sapphire
amplitude variations of up to 4\% over a few minutes.

A second large source of error is the non-ideal nature of the
beamsplitters. The calibrated birefringence properties are given
in \cite{clarke00a}.   It is sufficient here to state that
when purely horizontal light is input into PSB3, a power
leakage of approximately 0.9\% towards PBS4 was measured. Even
this small amount of reflected light can have large effects on the
ratio of light reaching detectors PD1 and PD2 due to interference
effects.

The drift of the interferometer during measurements was negligible
compared to the above errors.  This was evaluated by monitoring
the level of destructive interference that could be observed
on detectors PD1 and PD2 over several hours using high
light intensities.

The results in Fig. \ref{fig:trine-results} demonstrate that the
trine and antitrine measurements are in close agreement with
theoretical predictions.  The ratios
$\frac{2}{3}:\frac{1}{6}:\frac{1}{6}$ and
$\frac{1}{2}:\frac{1}{2}:0$ are clearly visible for the respective
measurements.  The RMS deviation of the trine and antitrine
results is 3.8\% from the theoretically expected values.  Of
specific importance are the antitrine measurements which are
theoretically expected to be zero.  The experimental measurements
are indeed very close to zero, with an average value of 1.6\%.
Fig. \ref{fig:rmsdev} also shows that the minimum RMS error from
the optimum theoretical values was obtained when the waveplate WP5
was at the angle corresponding to the theoretical value, within
the limits imposed by the small number of data points.

Close inspection of Fig. \ref{fig:trine-results} demonstrates
the effect of the leakage of PBS3.  For antitrine input state
$|\psi^3_3\rangle$
the light on PD1 and PD2 is split into the ratio 46:54.  In an
ideal experiment there would be no light traveling towards PBS4
from PBS2 and the split would be 50:50. Indeed, this ratio is
observed to better than $\pm 0.2\%$ if an opaque card is placed
between PBS2 and PBS4.  The leakage from PBS2, when interfering
with the light from PBS3, is enough to skew the results by 8\%.
This demonstrates the sensitivity of the apparatus to practical
sources of error.

\subsection{Tetrad Experiment}
\label{sec:tetrad_expt}

The first beamsplitter PBS1 in the interferometer was changed to
a non-polarizing beamsplitter.  Alignment of the interferometer
was achieved by simulating a conventional Mach-Zehnder operation
with vertically polarized input light.  Waveplate WP5 was aligned
with the slow axis in the vertical direction and WP7, a quarter
waveplate, rotated so that there was complete mixing of the output
states of the interferometer.

After the alignment, waveplate WP5 was rotated clockwise by
$54.74^{\circ}/2=27.37^{\circ}$. This corresponds to setting the fast axis
at $125.26^{\circ}/2=54.747^{\circ}$ anti-clockwise from the horizontal.
The angle between
two tetrad states is $125.26^{\circ}$ on the Poincar\'{e} sphere, or
$54.74^{\circ}$ in polarization.

The four tetrad and four antitetrad polarization states required
to perform this experiment were constructed using waveplates WP2-4
as described in Appendix \ref{app:jones}.  We were able to change
between the 8 input states in a matter of seconds using the
discrete predefined angles in our waveplate holders.

\begin{figure}
    \begin{minipage}{\columnwidth}
        \resizebox{\columnwidth}{!}{\includegraphics{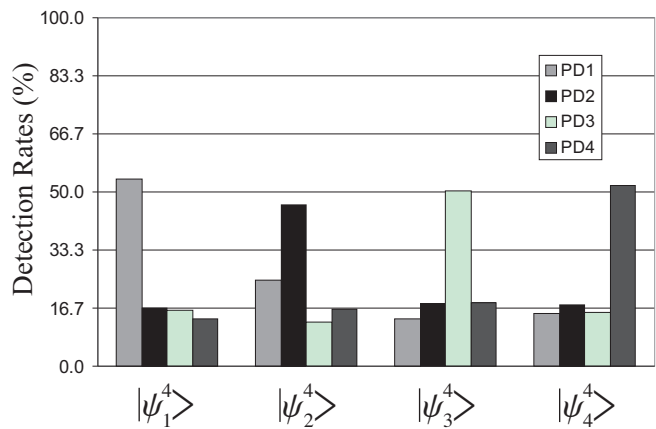}}
    \end{minipage}
    \hfill
    \caption{Histogram showing the tetrad results, normalized so that
    the sum of the counts in detectors 1 to 4 is $100\%$ for each input state.
    The theoretical ratios are
    $\frac{1}{2}:\frac{1}{6}:\frac{1}{6}:\frac{1}{6}$.}
    \label{fig:tetrad-res}
\end{figure}

\begin{figure}
    \begin{minipage}{\columnwidth}
        \resizebox{\columnwidth}{!}{\includegraphics{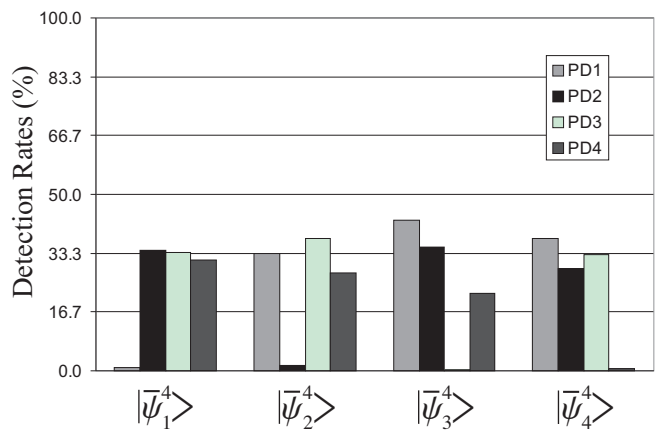}}
    \end{minipage}
    \hfill
    \caption{Histogram showing the antitetrad results, normalized so that
    the sum of the counts in detectors 1 to 4 is $100\%$ for each input state.
    The theoretical ratios are $\frac{1}{3}:\frac{1}{3}:\frac{1}{3}:0$.}
    \label{fig:Antitetrad}
\end{figure}

The experiment was performed by measuring the signal on detectors
PD1-4 as the input states were changed between the 8 states.  The
results of the tetrad and antitetrad measurements are shown in
Figs. \ref{fig:tetrad-res} and \ref{fig:Antitetrad}
after calibration and normalization.


As with the trine measurements, it is clear that the experimental
results are in good agreement with the theoretical predictions. The
$\frac{1}{2}:\frac{1}{6}:\frac{1}{6}:\frac{1}{6}$ and
$\frac{1}{3}:\frac{1}{3}:\frac{1}{3}:0$ ratios for the tetrad and
antitetrad states respectively are clearly evident.  The overall
RMS deviation from the theoretical values is 2.9\%.  Attention is
drawn to the antitetrad states where the count rate is
theoretically zero.  The average count rate in these four
measurements is 0.9\%.  This low count rate is very significant
for the mutual information that can be obtained, see next subsection.

\subsection{Experimental Mutual Information}
\label{sec:mutual-info}

The mutual information for both experiments was calculated in bits (log$_2$) using
Eq. (\ref{I(X:Y)eq}).   The conditional probabilities were obtained
from the experiment.  The values for $p_k$ were set to $\frac{1}{3}$
and $\frac{1}{4}$ for the trine and tetrad cases respectively,
corresponding to equal prior probabilities for the input states.
The results, summarized in Table \ref{tab:MIexpt},
are in excellent agreement with the
theoretical predictions using the model for detector noise
described in Appendix \ref{app:error}.  The mutual
information of both sets of antistates is significantly greater than the best
possible von Neumann measurement.

\renewcommand{\arraystretch}{1.2}	
\begin{table*}
\begin{tabular*}{\textwidth}{l@{\extracolsep{\fill}}cccc}
States & Experiment & Ideal Theory & Noisy Theory &  von Neumann
\\
\hline 
trine & 0.312 $^{+0.017}_{-0.08}$ & 0.333 & 0.302  & 0.459
\\
antitrine &  0.491 $^{+0.011}_{-0.027}$ & 0.585 & 0.486 & 0.459
\\
tetrad & 0.209 $^{+0.013}_{-0.010}$ & 0.208 & 0.194  & 0.311
\\
antitetrad &  0.363 $^{+0.09}_{-0.024}$& 0.415 & 0.355 & 0.311
\\
\hline
\end{tabular*}
\caption{Mutual information expressed in bits for the trine and
	tetrad states and their antistates.
	The values indicate the mutual information
	obtained in the experiment, in an ideal experiment, in an ideal
	experiment with detector noise analyzed using the error model
	described in Appendix B, and the
	maximum allowed mutual information in a von Neumann measurement.}
\label{tab:MIexpt}
\end{table*}
\renewcommand{\arraystretch}{1.0}

Experimental errors were estimated in a Monte Carlo simulation, assuming a flat $\pm$2.5\%
error distribution in the measurements.  The theoretical sensitivity of
the mutual information to measurement error was estimated
using a simple error model, see Appendix \ref{app:error}.
The parameter, $\Gamma$, was used to quantify the level of noise.
The average signal of the near-zero measurements in the trine and
tetrad experiments, 1.6\% and 0.9\% respectively,
was used to obtain the parameter $\Gamma=0.952$ and $0.964$ respectively.

In the antitrine and antitetrad cases it was found
that the sensitivity of the mutual information obtained is almost
entirely determined by the noise-induced counts at the detector that
would, in an ideal experiment, register no counts.
This is a consequence of the logarithmic nature of the mutual information.
In an ideal experiment, a count in detector PD$j$ means that state
$|\bar{\psi}^N_j\rangle$ was definitely not the input state.  The knowledge of
the input state has therefore been increased by a large
amount.  Any deviation from theoretically zero count rates in the
experiment will lead to a rapid increase in the errors associated
with this deduction.  Hence, the mutual information will decrease rapidly.
Nevertheless, in our experiments, the mutual information obtained
for both the antitrine and antitetrad states is still significantly higher than
the best possible von Neumann measurement.
To our knowledge this is first time this has been achieved
experimentally.
Measurement errors due to noise are the limiting factor
determining the mutual information that can be achieved experimentally.
The largest source of noise is due to ground
loop noise in the detectors.
The use of single photon
detectors gated with the pulses of light should reduce this noise
considerably and yield further increases in the mutual information obtained.

\section{Conclusions}
\label{sec:conc}

In quantum communications the possible signal states can be nonorthogonal.
This leads to problems in recovering the signal by detection.  In general,
we must accept the presence of errors.  We have realized detection
strategies that either minimise the error probability or maximise the mutual
information for the trine and tetrad states.
The mutual information is an important parameter in
communications, quantifying the maximum amount of information that
can be extracted from a signal.  In the context of measuring the
polarization of light, probability operator measures provide a
means of attaining higher mutual information values than von
Neuman measurements.  We have presented two separate experiments
using heavily attenuated pulses of light to verify the optimal POM on
trine and tetrad polarization states of light.  We have
demonstrated for the first time that the mutual information
obtainable using POMs can be significantly higher than the maximum
possible mutual information in a von Neuman measurement.
This illustrates the principle that generalised measurements can be of
greater utility than their more familiar von Neuman counterparts.

Using a flexible free space interferometer we were able to vary
all aspects of the measurement process deterministically. The
mutual information is extremely sensitive to the noise of the
detectors in which very low counts rates are expected.  We have
demonstrated that practical communications using POMs are
feasible.

\begin{acknowledgments}
This work was funded by the UK Engineering and Physical Sciences
Research Council grant numbers GR/L55216, GR/M60712 and GR/N17393.
AC, MS and SMB thank the British Council for financial support.
\end{acknowledgments}

\appendix

\section{Making an arbitrary polarization state}
\label{app:jones}

The convention used for the orientation of the waveplates is as follows.
Viewed in the direction of propagation, the fast axis of the waveplate
makes an angle $\phi/2$ with the horizontal in an anti-clockwise direction,
and the slow axis similarly makes an angle $\phi/2$ with the vertical.
Expressed in Jones matrix notation \cite{hecht87a},
the action of a half waveplate is thus
\begin{equation}
\Lambda_2(\phi/2)\left(\begin{array}{c} h \\ v \\ \end{array}\right)
= \left(\begin{array}{cc}
    \cos\phi &  \sin\phi \\
    \sin\phi & -\cos\phi \\
    \end{array}\right)
\left(\begin{array}{c} h \\ v \\ \end{array}\right),
\label{eq:halfplate}
\end{equation}
where $h$ and $v$ are the amplitudes of horizontal and vertical polarization
in the state respectively.
Similarly, the action of a quarter waveplate with its fast axis 
at an angle $\theta/2$ to the horizontal is
\renewcommand{\arraystretch}{1.2}
\begin{eqnarray}
&&\Lambda_4(\theta/2)\left(\begin{array}{c} h \\ v \\ \end{array}\right) \nonumber \\*
&&= \left(\begin{array}{cc}
    \sin^2(\theta/2) + i\cos^2(\theta/2) & \frac{i-1}{2}\sin\theta \\
    \frac{i-1}{2}\sin\theta & \cos^2(\theta/2) + i\sin^2(\theta/2) \\
    \end{array}\right)
\left(\begin{array}{c} h \\ v \\ \end{array}\right) \nonumber \\
&&=  \frac{1}{\sqrt{2}}\left(\begin{array}{cc}
        \cos\theta - i &  \sin\theta     \\
        \sin\theta     & -\cos\theta - i \\
        \end{array}\right)
\left(\begin{array}{c} h \\ v \\ \end{array}\right),
\label{eq:quarterplate}
\end{eqnarray}
\renewcommand{\arraystretch}{1.0}
where the overall phase of $e^{i3\pi/4}$ has been dropped.

\begin{figure}
    \begin{minipage}{\columnwidth}
    \begin{center}
        \resizebox{\columnwidth}{!}{\includegraphics{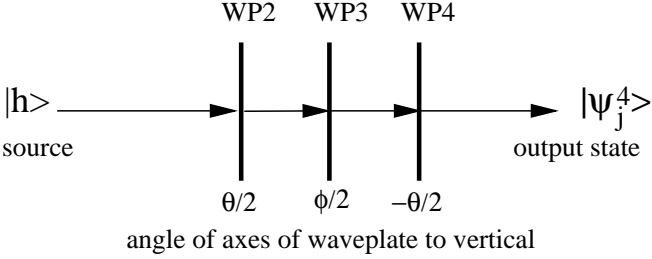}}
    \end{center}
    \end{minipage}
    \hfill
    \caption{Optical arrangement for making an arbitrary state.
    Waveplates WP2 and WP4 are quarter waveplates and WP3 is a half waveplate.}
    \label{fig:state}
\end{figure}

As can easily be verified using the above expressions,
an arbitrary polarization state can be made from horizontally
polarized input light using a sequence of three waveplates,
quarter-half-quarter, at the appropriate angles, see Fig. \ref{fig:state}.
If the input state is $|h\rangle$ (horizontal polarization), and
the orthogonal polarization is $|v\rangle$, the output state, $|\psi\rangle$,
is given by
\begin{equation}
|\psi\rangle = \cos\beta\,|h\rangle + \mathrm{e}^{i\gamma}\sin\beta\,|v\rangle,
\end{equation}
where $\sin\phi = -\sin\beta\,\sin\gamma$ and
$\tan\theta = \tan\beta\,\cos\gamma$,
with $\theta$, $\phi$ specifying the orientation of the three waveplates
as given in Fig.\ref{fig:state}.

\section{Details of tetrad optical implementation}
\label{app:tetrads}

Starting from the input states given by Eq. (\ref{eq:tetrads}),
at the stage indicated by the line AA on Fig. \ref{fig:tetrad},
we have
\begin{eqnarray}
|\psi^4_1\rangle_{\mathrm A} &=& \frac{1}{\sqrt2}\left(|\psi^4_1\rangle_{\mathrm L} + i|\psi^4_2\rangle_{\mathrm U}\right), \nonumber \\
|\psi^4_2\rangle_{\mathrm A} &=& \frac{1}{\sqrt2}\left(|\psi^4_2\rangle_{\mathrm L} - i|\psi^4_1\rangle_{\mathrm U}\right), \nonumber \\
|\psi^4_3\rangle_{\mathrm A} &=& \frac{1}{\sqrt2}\left(|\psi^4_3\rangle_{\mathrm L} + |\psi^4_4\rangle_{\mathrm U}\right), \nonumber \\
|\psi^4_4\rangle_{\mathrm A} &=& \frac{1}{\sqrt2}\left(|\psi^4_4\rangle_{\mathrm L} + |\psi^4_3\rangle_{\mathrm U}\right).
\end{eqnarray}
In terms of $|h\rangle$ and $|v\rangle$,
\begin{eqnarray}
|\psi^4_1\rangle_{\mathrm A} &=& \frac{1}{\sqrt6}\left(-|h_{\mathrm L}\rangle
        + \sqrt{2}\mathrm{e}^{-2\pi i/3}|v_{\mathrm L}\rangle
        - i|h_{\mathrm U}\rangle\right. \nonumber \\*
        &&+ i\sqrt{2}\mathrm{e}^{2\pi i/3}|v_{\mathrm U}\rangle\left.\mbox{\rule[-0.2em]{0em}{1.2em}}\right), \nonumber \\
|\psi^4_2\rangle_{\mathrm A} &=& \frac{1}{\sqrt6}\left(-|h_{\mathrm L}\rangle
        + \sqrt{2}\mathrm{e}^{2\pi i/3}|v_{\mathrm L}\rangle
        + i|h_{\mathrm U}\rangle\right. \nonumber \\*
        &&- i\sqrt{2}\mathrm{e}^{-2\pi i/3}|v_{\mathrm U}\rangle\left.\mbox{\rule[-0.2em]{0em}{1.2em}}\right), \nonumber \\
|\psi^4_3\rangle_{\mathrm A} &=& \frac{1}{\sqrt6}\left(-|h_{\mathrm L}\rangle
        + \sqrt{2}|v_{\mathrm L}\rangle
        + \sqrt{3}|h_{\mathrm U}\rangle\right), \nonumber \\
|\psi^4_4\rangle_{\mathrm A} &=& \frac{1}{\sqrt6}\left(\sqrt{3}|h_{\mathrm L}\rangle
        - |h_{\mathrm U}\rangle
        + \sqrt{2}|v_{\mathrm U}\rangle\right).
\end{eqnarray}
Then the $|h_{\mathrm L}\rangle$ components can reach photo-detector PD4
and the $|h_{\mathrm U}\rangle$ components can reach photo-detector PD3,
while the $|v_{\mathrm L}\rangle$ and $|v_{\mathrm U}\rangle$ components
continue round the lower and upper arms of the interferometer respectively.
Applying WP6 to $|v_{\mathrm L}\rangle$, converting it to
$|h_{\mathrm L}\rangle$, the states at the stage indicated by the line BB
in Fig. \ref{fig:tetrad} are
\begin{eqnarray}
|\psi^4_1\rangle_{\mathrm B} &=& \frac{1}{\sqrt6}\left(
          \sqrt{2}\mathrm{e}^{-2\pi i/3}|h_{\mathrm L}\rangle
        + i\sqrt{2}\mathrm{e}^{2\pi i/3}|v_{\mathrm U}\rangle\right. \nonumber \\*
        &&- i|PD3\rangle - |PD4\rangle\left.\mbox{\rule[-0.2em]{0em}{1.2em}}\right), \nonumber \\
|\psi^4_2\rangle_{\mathrm B} &=& \frac{1}{\sqrt6}\left(
          \sqrt{2}\mathrm{e}^{2\pi i/3}|h_{\mathrm L}\rangle
        - i\sqrt{2}\mathrm{e}^{-2\pi i/3}|v_{\mathrm U}\rangle\right. \nonumber \\*
        &&+ i|PD3\rangle - |PD4\rangle\left.\mbox{\rule[-0.2em]{0em}{1.2em}}\right), \nonumber \\
|\psi^4_3\rangle_{\mathrm B} &=& \frac{1}{\sqrt6}\left(
          \sqrt{2}|h_{\mathrm L}\rangle
        + \sqrt{3}|PD3\rangle - |PD4\rangle\right), \nonumber \\
|\psi^4_4\rangle_{\mathrm B} &=& \frac{1}{\sqrt6}\left(
          \sqrt{2}|v_{\mathrm U}\rangle
        - |PD3\rangle + \sqrt{3}|PD4\rangle\right).
\end{eqnarray}

Finally, the maximal mixing effect of WP9 brings the states into
\begin{eqnarray}
|\psi^4_1\rangle_{\mathrm C} &=& \frac{1}{\sqrt6}\left(
          -i\sqrt{3}|h\rangle - i|v\rangle
        - i|PD3\rangle - |PD4\rangle\right), \nonumber \\
|\psi^4_2\rangle_{\mathrm C} &=& \frac{1}{\sqrt6}\left(
        - |h\rangle - \sqrt{3}|v\rangle
        + i|PD3\rangle - |PD4\rangle\right), \nonumber \\
|\psi^4_3\rangle_{\mathrm C} &=& \frac{1}{\sqrt6}\left(
          |h\rangle + i|v\rangle
        + \sqrt{3}|PD3\rangle - |PD4\rangle\right), \nonumber \\
|\psi^4_4\rangle_{\mathrm C} &=& \frac{1}{\sqrt6}\left(
          i|h\rangle + |v\rangle
        - |PD3\rangle + \sqrt{3}|PD4\rangle\right),
\end{eqnarray}
from which the final outcome is easily seen to be
\begin{eqnarray}
|\psi^4_1\rangle_{\mathrm F} &=& -\frac{i}{\sqrt2}|PD1\rangle
        - \frac{i}{\sqrt6}|PD2\rangle \nonumber \\*
        &&+ \frac{i}{\sqrt6}|PD3\rangle
        - \frac{1}{\sqrt6}|PD4\rangle, \nonumber \\
|\psi^4_2\rangle_{\mathrm F} &=& -\frac{1}{\sqrt6}|PD1\rangle
        - \frac{1}{\sqrt2}|PD2\rangle \nonumber \\*
        &&- \frac{i}{\sqrt6}|PD3\rangle
        - \frac{1}{\sqrt6}|PD4\rangle, \nonumber \\
|\psi^4_3\rangle_{\mathrm F} &=& \frac{1}{\sqrt6}|PD1\rangle
        + \frac{i}{\sqrt6}|PD2\rangle \nonumber \\*
        &&+ \frac{1}{\sqrt2}|PD3\rangle
        - \frac{1}{\sqrt6}|PD4\rangle, \nonumber \\
|\psi^4_4\rangle_{\mathrm F} &=& \frac{i}{\sqrt6}|PD1\rangle
        + \frac{1}{\sqrt6}|PD2\rangle \nonumber \\*
        &&- \frac{1}{\sqrt6}|PD3\rangle
        + \frac{1}{\sqrt2}|PD4\rangle,
\label{eq:tetrad-final}
\end{eqnarray}
where the amplitudes give the probabilities of
each photodetector finding the photon.

\section{Mutual Information Error Model}
\label{app:error}

The mutual information obtained is very sensitive to noise.  In
this appendix we present a simple model for detector noise and use
it to analyze our measurements of mutual information.
Rather than attempt to account for all sources of noise, our model simply
assumes that the noise randomly distributes a proportion of the detection
events among all the detectors.

The probability for detecting a photon in detector $j$ (outcome $y_j$) is
associated with the POM element $\Pi_j$.  Noise can cause a
photocurrent to occur in a detector in the absence of any photons.
Such noise-inducing
events are randomly distributed and so we can modify our POM
elements to account for these noisy events by replacing POM
elements $\{\Pi_j \}$ for an ideal measurement with
\begin{equation}
\Pi_j^{\mathrm{noise}}=\Gamma\Pi_j + \frac{1-\Gamma}{N}.
\end{equation}
Here $N$ is the number of measurement outcomes and $\Gamma$ is a
positive number ($\Gamma\le 1$) that characterizes
the noise.  For $\Gamma=1$ there is no noise, but for $\Gamma=0$,
noise accounts for all detection events.  The conditional
probabilities in the presence of noise become
\begin{equation}
P(y_j\vert\psi^N_k)=\langle\psi^N_k|\Pi_j^{\mathrm{noise}}|\psi^N_k\rangle.
\end{equation}

Given a channel in which the input states $|\psi^N_k\rangle$
occur with equal probability we find that the mutual information
becomes
\begin{eqnarray}
\label{mutualinfotrine}
I(\{|\psi^N_k\rangle\}:&&\{y_j\})=\left(\frac{1+\Gamma}{N}\right)\log(1+\Gamma) \nonumber \\*
      +&&\left(1-\frac{1+\Gamma}{N}\right)\log\left(1-\frac{\Gamma}{N-1}\right).
\end{eqnarray}
If, instead, the antistates $|\bar{\psi}^N_k\rangle$ are used as input, the
mutual information is
\begin{eqnarray}
I(\{|\bar{\psi}^N_k\rangle\}:&&\{y_j\})=\left(\frac{1-\Gamma}{N}\right)\log(1-\Gamma) \nonumber \\*
      +&&\left(1-\frac{1-\Gamma}{N}\right)\log\left(1+\frac{\Gamma}{N-1}\right).
\label{mutualinfoantitrine}
\end{eqnarray}
It is interesting to note that \ref{mutualinfotrine} and
\ref{mutualinfoantitrine} differ in form only through the sign of
$\Gamma$.

We can estimate the value of $\Gamma$ by considering the
probability for the measurement of the
antistate $|\bar{\psi}^N_k\rangle$ to give the result $y_k$, 
corresponding to ``not-$|\bar{\psi}^N_k\rangle$''.
This is forbidden for
an ideal, noiseless experiment but will occur with probability
$(1-\Gamma)/N$ in the presence of noise.  Comparing this with
the experimentally observed values provides us with a value for
$\Gamma$.

\begin{figure}
    \begin{minipage}{\columnwidth}
        \resizebox{\columnwidth}{!}{\includegraphics{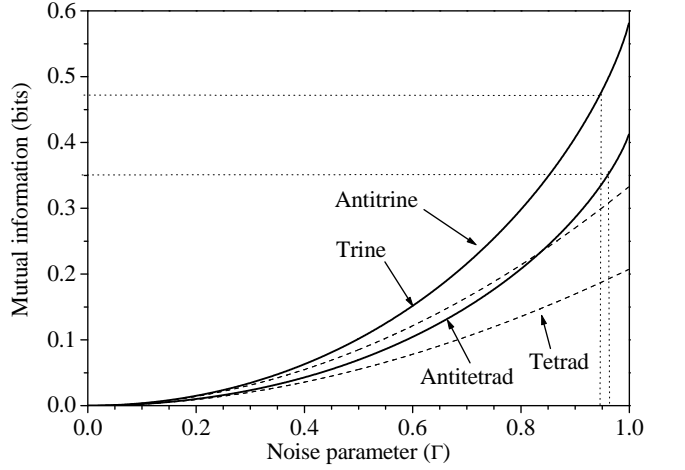}}
    \end{minipage}
    \hfill
    \caption{The theoretical variation of the mutual information with the
    detector noise, given by parameter $\Gamma$, for the trine,
    antitrine, tetrad and antitetrad states.  Included are
    read-off lines to show the theoretical values of the mutual information
    for the antitrine and antitetrad measurements, given the experimental
    values of $\Gamma=0.952$ and $0.964$ respectively.}
    \label{fig:appbfig}
\end{figure}
Figure \ref{fig:appbfig} shows the variation of the mutual
information obtained as $\Gamma$ is varied.  The sensitivity to
noise in the antitrine and antitetrad states increases markedly
as $\Gamma$ tends to 1. This demonstrates the sensitivity of the
mutual information to noise in the detector which, in an ideal
experiment, would register no counts.  The mutual information of
trine and tetrad measurements does not increase as rapidly because
there are no detectors in which zero counts are expected.


\bibliography{dis}

\begin{thebibliography}{10}
\expandafter\ifx\csname bibnamefont\endcsname\relax
  \def\bibnamefont#1{#1}\fi
\expandafter\ifx\csname bibfnamefont\endcsname\relax
  \def\bibfnamefont#1{#1}\fi
\expandafter\ifx\csname url\endcsname\relax
  \def\url#1{\texttt{#1}}\fi
\expandafter\ifx\csname urlprefix\endcsname\relax\def\urlprefix{URL }\fi
\providecommand{\bibinfo}[2]{#2}
\providecommand{\eprint}[2][]{\url{#2}}

\bibitem{phoenix95a}
\bibinfo{author}{\bibfnamefont{S.~J.~D.} \bibnamefont{Phoenix}}
  \bibnamefont{and} \bibinfo{author}{\bibfnamefont{P.~D.}
  \bibnamefont{Townsend}}, \bibinfo{journal}{Contemp. Phys.}
  \textbf{\bibinfo{volume}{36}}, \bibinfo{pages}{165} (\bibinfo{year}{1995}),
  \bibinfo{note}{and references therein}.

\bibitem{fuchs97a}
\bibinfo{author}{\bibfnamefont{C.~F.} \bibnamefont{Fuchs}},
  \bibinfo{journal}{Phys. Rev. Lett.}
  \textbf{\bibinfo{volume}{79}}(\bibinfo{number}{6}), \bibinfo{pages}{1162}
  (\bibinfo{year}{1997}).

\bibitem{helstrom76a}
\bibinfo{author}{\bibfnamefont{C.~W.} \bibnamefont{Helstrom}},
  \emph{\bibinfo{title}{Quantum Detection and Estimation Theory}}
  (\bibinfo{publisher}{Academic Press, New York}, \bibinfo{address}{New York},
  \bibinfo{year}{1976}).

\bibitem{peres93a}
\bibinfo{author}{\bibfnamefont{A.}~\bibnamefont{Peres}},
  \emph{\bibinfo{title}{Quantum Theory: Concepts and Methods}}
  (\bibinfo{publisher}{Kluwer Academic Publishers, Dordrecht},
  \bibinfo{address}{Dordrecht}, \bibinfo{year}{1993}), \bibinfo{note}{also
  called a positive operator-valued measure}.

\bibitem{holevo73b}
\bibinfo{author}{\bibfnamefont{A.~S.} \bibnamefont{Holevo}},
  \bibinfo{journal}{Problemy Peredachi Informatsii}
  \textbf{\bibinfo{volume}{9}}, \bibinfo{pages}{31} (\bibinfo{year}{1973}),
  \bibinfo{note}{{P}roblems of {I}nformation {T}ransmision (USSR), \textbf{9},
  110}.

\bibitem{sasaki96a}
\bibinfo{author}{\bibfnamefont{M.}~\bibnamefont{Sasaki}} \bibnamefont{and}
  \bibinfo{author}{\bibfnamefont{O.}~\bibnamefont{Hirota}},
  \bibinfo{journal}{Phys. Rev. A}
  \textbf{\bibinfo{volume}{54}}(\bibinfo{number}{4}), \bibinfo{pages}{2728}
  (\bibinfo{year}{1996}).

\bibitem{barnett97a}
\bibinfo{author}{\bibfnamefont{S.~M.} \bibnamefont{Barnett}} \bibnamefont{and}
  \bibinfo{author}{\bibfnamefont{E.}~\bibnamefont{Riis}}, \bibinfo{journal}{J.
  Mod. Opt.} \textbf{\bibinfo{volume}{44}}, \bibinfo{pages}{1061}
  (\bibinfo{year}{1997}).

\bibitem{ivanovic87a}
\bibinfo{author}{\bibfnamefont{I.~D.} \bibnamefont{Ivanovic}},
  \bibinfo{journal}{Phys. Lett. A} \textbf{\bibinfo{volume}{123}},
  \bibinfo{pages}{257} (\bibinfo{year}{1987}).

\bibitem{dieks88a}
\bibinfo{author}{\bibfnamefont{D.}~\bibnamefont{Dieks}},
  \bibinfo{journal}{Phys. Lett. A} \textbf{\bibinfo{volume}{126}},
  \bibinfo{pages}{303} (\bibinfo{year}{1988}).

\bibitem{peres88a}
\bibinfo{author}{\bibfnamefont{A.}~\bibnamefont{Peres}},
  \bibinfo{journal}{Phys. Lett. A} \textbf{\bibinfo{volume}{128}},
  \bibinfo{pages}{19} (\bibinfo{year}{1988}).

\bibitem{jaeger95a}
\bibinfo{author}{\bibfnamefont{G.}~\bibnamefont{Jaeger}} \bibnamefont{and}
  \bibinfo{author}{\bibfnamefont{A.}~\bibnamefont{Shimony}},
  \bibinfo{journal}{Phys. Lett. A} \textbf{\bibinfo{volume}{197}},
  \bibinfo{pages}{83} (\bibinfo{year}{1995}).

\bibitem{chefles98a}
\bibinfo{author}{\bibfnamefont{A.}~\bibnamefont{Chefles}} \bibnamefont{and}
  \bibinfo{author}{\bibfnamefont{S.~M.} \bibnamefont{Barnett}},
  \bibinfo{journal}{J. Mod. Opt.} \textbf{\bibinfo{volume}{45}},
  \bibinfo{pages}{1295} (\bibinfo{year}{1998}).

\bibitem{huttner96a}
\bibinfo{author}{\bibfnamefont{B.}~\bibnamefont{Huttner}},
  \bibinfo{author}{\bibfnamefont{A.}~\bibnamefont{Muller}},
  \bibinfo{author}{\bibfnamefont{J.~D.} \bibnamefont{Gautier}},
  \bibinfo{author}{\bibfnamefont{H.}~\bibnamefont{Zbinden}}, \bibnamefont{and}
  \bibinfo{author}{\bibfnamefont{N.}~\bibnamefont{Gisin}},
  \bibinfo{journal}{Phys. Rev. A} \textbf{\bibinfo{volume}{54}},
  \bibinfo{pages}{3783} (\bibinfo{year}{1996}).

\bibitem{clarke00a}
\bibinfo{author}{\bibfnamefont{R.~B.~M.} \bibnamefont{Clarke}},
  \bibinfo{author}{\bibfnamefont{A.}~\bibnamefont{Chefles}},
  \bibinfo{author}{\bibfnamefont{S.~M.} \bibnamefont{Barnett}},
  \bibnamefont{and} \bibinfo{author}{\bibfnamefont{E.}~\bibnamefont{Riis}},
  \bibinfo{journal}{Phys. Rev. A}  (\bibinfo{year}{2001}),
  \bibinfo{note}{accepted February 2001}.

\bibitem{peres92a}
\bibinfo{author}{\bibfnamefont{A.}~\bibnamefont{Peres}} \bibnamefont{and}
  \bibinfo{author}{\bibfnamefont{W.}~\bibnamefont{Wootters}},
  \bibinfo{journal}{Phys. Rev. Lett.} \textbf{\bibinfo{volume}{66}},
  \bibinfo{pages}{1119} (\bibinfo{year}{1992}).

\bibitem{hausladen94a}
\bibinfo{author}{\bibfnamefont{P.}~\bibnamefont{Hausladen}} \bibnamefont{and}
  \bibinfo{author}{\bibfnamefont{W.~K.} \bibnamefont{Wootters}},
  \bibinfo{journal}{J. Mod. Optics}
  \textbf{\bibinfo{volume}{41}}(\bibinfo{number}{12}), \bibinfo{pages}{2385}
  (\bibinfo{year}{1994}).

\bibitem{born87a}
\bibinfo{author}{\bibfnamefont{M.}~\bibnamefont{Born}} \bibnamefont{and}
  \bibinfo{author}{\bibfnamefont{E.}~\bibnamefont{Wolf}},
  \emph{\bibinfo{title}{Principles of Optics}} (\bibinfo{publisher}{Pergamon
  Press, Oxford}, \bibinfo{year}{1987}), \bibinfo{note}{p.31}.

\bibitem{davies78a}
\bibinfo{author}{\bibfnamefont{E.~B.} \bibnamefont{Davies}},
  \bibinfo{journal}{IEEE Trans. Inform. Theory} \textbf{\bibinfo{volume}{24}},
  \bibinfo{pages}{596} (\bibinfo{year}{1978}).

\bibitem{holevo73a}
\bibinfo{author}{\bibfnamefont{A.~S.} \bibnamefont{Holevo}},
  \bibinfo{journal}{J. Multivariate Analysis} \textbf{\bibinfo{volume}{3}},
  \bibinfo{pages}{337} (\bibinfo{year}{1973}).

\bibitem{yuen75a}
\bibinfo{author}{\bibfnamefont{H.~P.} \bibnamefont{Yuen}},
  \bibinfo{author}{\bibfnamefont{R.~S.} \bibnamefont{Kennedy}},
  \bibnamefont{and} \bibinfo{author}{\bibfnamefont{M.}~\bibnamefont{Lax}},
  \bibinfo{journal}{IEEE Trans. Information Theory}
  \textbf{\bibinfo{volume}{21}}(\bibinfo{number}{2}), \bibinfo{pages}{125}
  (\bibinfo{year}{1975}).

\bibitem{phoenix00a}
\bibinfo{author}{\bibfnamefont{S.~J.~D.} \bibnamefont{Phoenix}},
  \bibinfo{author}{\bibfnamefont{S.~M.} \bibnamefont{Barnett}},
  \bibnamefont{and} \bibinfo{author}{\bibfnamefont{A.}~\bibnamefont{Chefles}},
  \bibinfo{journal}{J. Mod. Opt.} \textbf{\bibinfo{volume}{47}},
  \bibinfo{pages}{507} (\bibinfo{year}{2000}).

\bibitem{sasaki99a}
\bibinfo{author}{\bibfnamefont{M.}~\bibnamefont{Sasaki}},
  \bibinfo{author}{\bibfnamefont{S.~M.} \bibnamefont{Barnett}},
  \bibinfo{author}{\bibfnamefont{R.}~\bibnamefont{Jozsa}},
  \bibinfo{author}{\bibfnamefont{M.}~\bibnamefont{Osaki}}, \bibnamefont{and}
  \bibinfo{author}{\bibfnamefont{O.}~\bibnamefont{Hirota}},
  \bibinfo{journal}{Phys. Rev. A}
  \textbf{\bibinfo{volume}{59}}(\bibinfo{number}{5}), \bibinfo{pages}{3325}
  (\bibinfo{year}{1999}).

\bibitem{ban97a}
\bibinfo{author}{\bibfnamefont{M.}~\bibnamefont{Ban}},
  \bibinfo{author}{\bibfnamefont{K.}~\bibnamefont{Kurokawa}},
  \bibinfo{author}{\bibfnamefont{R.}~\bibnamefont{Momose}}, \bibnamefont{and}
  \bibinfo{author}{\bibfnamefont{O.}~\bibnamefont{Hirota}},
  \bibinfo{journal}{Int. J. Theor. Phys.}
  \textbf{\bibinfo{volume}{36}}(\bibinfo{number}{6}), \bibinfo{pages}{1269}
  (\bibinfo{year}{1997}).

\bibitem{cover91a}
\bibinfo{author}{\bibfnamefont{T.}~\bibnamefont{Cover}} \bibnamefont{and}
  \bibinfo{author}{\bibfnamefont{J.}~\bibnamefont{Thomas}},
  \emph{\bibinfo{title}{Elements of Information Theory}}
  (\bibinfo{publisher}{John Wiley and Sons}, \bibinfo{address}{New York},
  \bibinfo{year}{1991}).

\bibitem{osaki00a}
\bibinfo{author}{\bibfnamefont{M.}~\bibnamefont{Osaki}},
  \bibinfo{author}{\bibfnamefont{M.}~\bibnamefont{Ban}}, \bibnamefont{and}
  \bibinfo{author}{\bibfnamefont{O.}~\bibnamefont{Hirota}}, in
  \emph{\bibinfo{booktitle}{Proceedings of the 4th conference on Quantum
  Communication, Measurement, and Computating, Evanston, Illinois, USA (1998)}}
  (\bibinfo{publisher}{Plenum Press}, \bibinfo{address}{New York},
  \bibinfo{year}{2000}).

\bibitem{naimark40a}
\bibinfo{author}{\bibfnamefont{M.~A.} \bibnamefont{Naimark}},
  \bibinfo{journal}{Izv. Akad. Nauk. SSSR Ser. Mat.}
  \textbf{\bibinfo{volume}{4}}, \bibinfo{pages}{277} (\bibinfo{year}{1940}).

\bibitem{horowitz89a}
\bibinfo{author}{\bibfnamefont{P.}~\bibnamefont{Horowitz}} \bibnamefont{and}
  \bibinfo{author}{\bibfnamefont{W.}~\bibnamefont{Hill}},
  \emph{\bibinfo{title}{The Art of Electronics}} (\bibinfo{publisher}{Cambridge
  University Press}, \bibinfo{address}{Cambridge UK}, \bibinfo{year}{1989}).

\bibitem{hecht87a}
\bibinfo{author}{\bibfnamefont{E.}~\bibnamefont{Hecht}},
  \emph{\bibinfo{title}{Optics}} (\bibinfo{publisher}{Addison-Wesley, Reading,
  Mass}, \bibinfo{year}{1987}), \bibinfo{note}{2nd ed., (see, for example p.321
  ff)}.

\end{thebibliography}


\end{document}